\documentclass[openacc]{rstransa}%%%%where rstrans is the template name

\newcommand{\alfven}{Alfv{\'e}n\,}
\newcommand{\degrees}{^{\circ}}

\begin{document}

%%%% Article title to be placed here
\title{The emergence of magnetic flux and its role on the onset of solar dynamic events}

\author{%%%% Author details
V. Archontis$^{1}$, P. Syntelis$^{1}$}

%%%%%%%%% Insert author address here
\address{$^{1}$St Andrews University, Mathematics Institute, St Andrews KY16 9SS, UK}

%%%% Subject entries to be placed here %%%%
\subject{Solar physics}

%%%% Keyword entries to be placed here %%%%
\keywords{Sun, magnetic fields, jets, eruptions}

%%%% Insert corresponding author and its email address}
\corres{V. Archontis\\
\email{va11@st-andrews.ac.uk}}

%%%% Abstract text to be placed here %%%%%%%%%%%%
\begin{abstract}
A plethora of solar dynamic events, such as the formation of active regions, the emission of jets and the occurrence of eruptions is often associated to the emergence of magnetic flux from the interior of the Sun to the surface and above. Here, we present a short review on the onset, driving and/or triggering of such events by magnetic flux emergence. We briefly describe some key observational examples, theoretical aspects and numerical simulations, towards revealing the mechanisms that govern solar dynamics and activity related to flux emergence. We show that the combination of important physical processes like shearing and reconnection of magnetic fieldlines in emerging flux regions or at their vicinity, can power some of the most dynamic phenomena in the Sun on various temporal and spatial scales. Based on previous and recent observational and numerical studies, we highlight that, in most cases, none of these processes alone can drive and also trigger explosive phenomena releasing considerable amount of energy towards the outer solar atmosphere and space, such as flares, jets and large-scale eruptions (e.g. CMEs). In addition, one has to take into account the physical properties of the emerging field (e.g. strength, amount of flux, relative orientation to neighbouring and pre-existing magnetic fields, etc.) in order to better understand the exact role of magnetic flux emergence on the onset of solar dynamic events.       
\end{abstract}
%%%%%%%%%%%%%%%%%%%%%%%%%%%

%%%%%%%%%% Insert the texts which can accomodate on firstpage in the tag "fmtext" %%%%%
%\begin{fmtext}
\maketitle
\section{Introduction}
\begin{figure*}
\centering\includegraphics[width=0.9\textwidth]{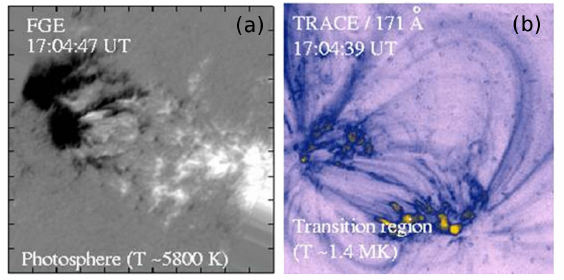}
\caption{Formation of an Active Region following magnetic flux emergence:
(a) vertical component of photospheric magnetic field (white is positive and black is negative
magnetic polarity). Small-scale mixed polarity magnetic field appears between the two sunspots.
(b) View of (a) in EUV 171\AA (transition region), showing how magnetic loops join the opposite polarity fields in the AR (adapted from \cite{2014SSRv}).
}
\label{fig:observations}
\end{figure*}

%%%% Insert A head here
One of the key processes related to the solar magnetic activity is the emergence of magnetic flux from the solar interior towards the visible surface of the Sun (photosphere) and the outer solar atmosphere. It is generally accepted that dynamo-generated magnetic fields are transported from the deep convection zone to the photopshere by magnetic buoyancy\cite{Parker_1955}. As the magnetic fields rise towards the solar surface, convective motions (updrafts and downdrafts) have an impact on the shape of the emerging fields, which may develop a serpentine-like configuration over a wide range of spatial scales. On small scales (e.g.  1-2 Mm), granular convection affects the emergence of magnetic fields, which appear to the photosphere in the form of small magnetic bipoles. On large scales (e.g. 100 Mm), the interplay between convection, photospheric motions and interaction of the small emerging bipoles (cancellation, coalescence etc.) can lead to the formation of sunspots and Active Regions (ARs) \cite{Zwaan_1985} (Fig.~\ref{fig:observations}). Observations have shown that explosive phenomena, such as jets, flares and eruptions (e.g. Coronal Mass Ejections; CMEs) often occur in ARs. In the past, a series of review papers have summarized: the evolution of ARs from their emergence through their decay \cite{2015LRSP}, the emergence of magnetic flux along the solar cycle \cite{2014SSRv}, the origin and evolution of solar eruptions in connection to flux emergence \cite{2018SSRv}, theoretical aspects and dynamics of multi-scale flux emergence\cite{Cheung_etal2014}, the nature of magnetic flux emergence and the associated magnetic activity in 3D numerical simulations \cite{2008JGRA} and the physical mechanism(s) of eruptions and CMEs \cite{2012AdSpR}. 

In this review, we focus on the mechanisms of solar eruptions, which originate in emerging flux regions. 
Firstly, we present some observational examples of eruptive events associated with emerging/emerged magnetic flux. Secondly, we discuss the physical mechanisms, which might be responsible for the onset of eruptions and eruption-driven events (e.g. jets). We also present some of the most recent developments and advances related to numerical simulations of magnetic flux emergence leading to eruptions. We mainly focus on numerical models showing the formation mechanism of magnetic flux ropes and their eruptivity, which can evolve into large-scale eruptions (e.g. CMEs) or smaller-scale ejections of hot and cool plasma into the outer solar atmosphere.
%\end{fmtext}

\maketitle
\section{Observations and Theory}
The eruption of solar filaments triggering CMEs is an explosive phenomenon, which is often associated with flaring activity and the destabilization of the coronal magnetic field. Various observational studies have reported on the pre-eruptive stage of the eruption, the onset and the propagation of the erupting field towards the interplanetary space (e.g., \cite{Canou_Amari2010}, \cite{Vourlidas_etal2012,Zuccarello_etal2014,Chintzoglou_etal2015,Reeves_etal2015,Syntelis_etal2016,Patsourakos_etal2016}). It is believed that the core of the erupting field has the form of a sheared arcade or a twisted magnetic flux tube (i.e. magnetic flux rope, MFR) (e.g., \cite{Cheng_etal2011,Green_etal2011,Zhang_etal2012,Patsourakos_etal2013}). 
CMEs usually emanate from ARs, throughout their lifetimes, from their initial emerging phase to their decay phase (e.g. \cite{2002SoPh}). Their onset has also been associated with emergence of magnetic flux. On the one hand, it has been shown that during the initial emerging phase of an AR, emerging flux \textit{alone} has the necessary energy to produce eruptions (e.g. \cite{Demoulin_etal2002,2003Nindos}) by itself. On the other hand, in many occasions, emerging flux acts primarily as a trigger for the eruption of a pre-existing filament (\cite{Feynman_etal1995,Williams2005}).

% ============ FIGURE ============================
% ============ FIGURE ============================
\subsection{Observational examples}

An observational survey of eruptions \cite{Feynman_etal1995} reported that most of the eruptions studied, there was new flux emerging at the close neighborhood of a filament. They found that flux emergence started a few days before the ejective eruption of the filament, which indicates that the interaction between the emerging field and the pre-existing magnetic field of (or enveloping) the filament triggered the eruption. In fact, it was reported that the orientation between the two interacting magnetic flux systems was such that efficient reconnection between them occurred, leading to the onset of a CME. Moreover, it was found that eruptions were not observed in most of the cases where flux emergence did not occur nearby the filament. The authors stated that these results are indicative but not conclusive on whether flux emergence is a necessary condition for the onset of eruptions. 

Another observational study \cite{Zhang_etal2008} presented a statistical survey, comprising newly emerging and well-developed ARs and CME-source regions, towards understanding the relationship between surface magnetic field variation and CME initiation. By measuring the total magnetic flux and the flux variation rate, it was found that magnetic flux increases (decreases) before the CME initiation in 60 (40)\% of the CME-source regions. Also, small-scale magnetic flux emergence was observed to occur in 91\% of CME-source regions. In general, the normalized flux variation rate was found to be very similar in CME-source regions and ARs, and much smaller in newly emerging ARs. Based on their study, the authors concluded that flux emergence alone does not necessarily lead to the onset of CMEs.

The connection between emerging flux and eruptions was also reported in an observational study of an AR with intense flaring and eruptive activity (NOAA 10501) \cite{Chandra2010}. In this AR, new magnetic flux emerged with opposite sign (negative) of helicity compared to the pre-existing AR. After studying the amount of helicity, which was injected into the AR due to flux emergence prior to the onset of the eruptions, it was concluded that the triggering mechanism was intricate. It was the combination of: flux emergence, interaction of the new emerging bipole(s) with the pre-existing field and the shearing between the magnetic polarities of the two flux systems (emerging and pre-existing).

A similar study \cite{Nindos2002} reported on the role of flux emergence and shearing flows at the photosphere in injecting helicity into the corona before the onset of a CME (NOAA 9165). It was found that shear flows were not so effective in injecting helicity, flux emergence injected most of the magnetic helicity into the outer solar atmosphere compared to that ejected by the CME. Obviously, when there is no flux emergence or when it is not so profound, convection-driven photospheric motions and shearing may inject a considerable amount of helicity into the corona 
(e.g. \cite{Savcheva2012}), playing an important role to the onset of eruptions. However, it has been reported that the frequency rate for CMEs is higher when new flux is emerging at the solar surface (e.g. \cite{Green2003}).

Overall, during the pre-eruptive phase of a CME, energy is built-up in the corona driven by various processes such as flux emergence, photospheric motions and differential rotation. In this context, magnetically driven shearing along strong polarity inversion lines can provide free magnetic energy to the system (e.g. within an AR) large enough to power a CME. This process will be discussed more in the next chapter. In the following, we will focus more on theoretical aspects of the triggering mechanism(s) of CMEs and the respective role of flux emergence.

\subsection{Theoretical aspects}
Studies on solar eruptions have repeatedly shown/suggested that the core of the erupting structure (filament) 
has a flux-rope-like configuration (e.g. \cite{Green_etal2011,Patsourakos_etal2013} etc.). One mechanism, which can lead to the formation of 
a MFR, is the combination of shearing along strong polarity inversion lines and reconnection 
of the sheared fieldlines (\cite{vanBallegooijen_etal1989,Magara_etal2001,Archontis2008,Fan2009,Priest_etal2017}). The field above the MFR consists of magnetic fieldlines, which envelope the core of the 
erupting structure (i.e. envelope field). The overall system of the erupting field consists mainly of the MFR and the envelope field. In principle, the energy that is stored into the system is a crucial 
parameter that affects its stability. Once the system looses its stability, it may erupt in a confined or an ejective manner (e.g. \cite{Archontis_etal2012, Leake_etal2013a, Leake_etal2014}). In the first 
cases, the erupting field may remain confined in the low corona due to the presence of a strong overlying magnetic field. In the second case, the erupting field goes through a slow rise phase
(e.g. \cite{Sterling2005, Schrijver2008}), followed by a rapid rise
phase and, possibly, a propagation phase towards the outer space at approximately 
constant speed. The fast-rise phase is typically associated with acceleration 
and an exponential height-time profile of the erupting structure.  

There are two broad types of mechanisms that have been suggested for the driving and/or 
triggering of the solar eruptions. One is associated with the occurrence of an (ideal) 
instability and the other with the (non-ideal) process of magnetic reconnection.  

For the first type of mechanism, a crucial parameter that affects the eruption of a MFR
is how the strength of the envelope magnetic field, constraining the MFR, varies with height. This is related to the so-called torus instability (\cite{Bateman_1978, Kliem_etal2006}). These initial studies have shown that when the rate of decrease of the
envelope magnetic field exceeds a critical value ($n = 1.5$), the current-currying FR 
becomes unstable and it erupts. The rate of decrease is typically referred as the torus or decay index. Later studies \cite{Demoulin_etal2010,Zuccarello_etal2015} reported that the critical value of the torus
index depends on various parameters, such as the thickness of the axial current of the MFR (i.e. current channel). For instance, in a case of a thin straight (circular) current channel, the torus index was 
found to be equal to 1 (1.5). For wide channel, the torus index varies in the range 1.1 to 1.5, depending also on how much the MFR expands outwards (due to magnetic pressure) during its rising motion. Numerical simulations of magnetic flux emergence have shown that the torus index can take even higher values, e.g. to vary within the range 1.7 to more than 2 (e.g. \cite{Fang_etal2010, An_Magara_2013}). 

Another ideal instability, which could be responsible for the onset of 
solar eruptions is the so-called helical kink instability (\cite{Anzer_1968, Torok_etal2004}). This instability occurs when the twist of the MFR exceeds a
certain value and the axis of the MFR develops a helical shape. The critical 
value / threshold depends on various parameters, such as 
the aspect ratio of the rope, the geometrical shape of the original MFR (e.g., toroidal-like or cylindrical-like shaped MFR) and the line-tying effect (e.g., \cite{Hood_Priest_1981, Torok_etal2004}). 

For the second type of mechanism responsible for the drive and/or trigger of solar eruptions, magnetic reconnection seems to be the key process. Reconnection can reduce the downwards tension of the envelope field surrounding the MFR and lead to an eruption. There are two major ways through which this process is feasible: tether-cutting and breakout reconnection. 

During tether-cutting reconnection \cite{Moore1992}, the lower segments of the envelope 
fieldlines can reconnect along a polarity inversion line (PIL), at a vertical current sheet underneath the current-carrying MFR (e.g. \cite{Moore_etal2001}). 
In this way, reconnection ``cuts the tethers'' of the envelope fieldlines \cite{Sturrock1989} and, thus, it reduces the downward tension of the envelope field. 
As the MFR rises: i)
more magnetized plasma is dragged into the current sheet underneath it and ii) the surrounding and overlying fields are vertically stretched. 
Eventually, the highly stretched magnetic fields reconnect at the current sheet underneath the MFR. In turn, the fast reconnection that occurs there, helps the MFR to erupt in an ejective manner. The upward reconnection jet from the current sheet assists the ejective eruption of the MFR. The ejective eruption is achieved through this process due to an imbalance between the downward tension force of the envelope field and the upward tension force of the reconnected fieldlines at the current sheet. The whole process can become explosive, leading to the onset of flares underneath the erupting MFR, in a similar manner to the formation of flare loops in the wake of a CME. 3D numerical experiments of magnetic flux emergence have shown that the above process can drive and trigger CME-like eruptions (see next section). 

In the case of a breakout reconnection, the removal of the downward tension of the envelope field occurs via reconnection above a MFR or a sheared arcade \cite{Antiochos_etal1999}. It has been shown that this kind of reconnection can drive eruptions in an explosive manner under (at least) two conditions: i) reconnection at a null point above the erupting field does not start during the initial phase of energy build-up and ii) the reconnection rate is slow during the fast-rise phase of the eruption. It has also been shown \cite{DeVore2005} that the ratio of the fluxes above and below the null point affects the efficiency of this mechanism. The relative amount of fluxes should be enough to keep breakout reconnection at work during a considerable amount of time for an ejective eruption to occur. The two flux systems could belong to the envelope field and to a pre-existing ambient magnetic field. Therefore, another parameter which affects the efficiency of their reconnection is their relative orientation. For instance, if the two systems have an anti-parallel relative orientation when they come into contact, reconnection between them becomes very effective (e.g., \cite{Antiochos_etal1999, Archontis_etal2012, Karpen_etal2012, Leake_etal2014}). Also, the relative field strengths of the
interacting magnetic systems can affect the eruption. Various numerical studies of 
flux emergence have shown that the eruption could be ejective or confined
or the erupting MFR may experience annihilation through the
interaction with the pre-existing magnetic field (e.g., \cite{Galsgaard2007, Archontis_etal2012, Leake_etal2014, Jiang_etal2016}).

The removal or reduce of the downward tension of an envelope field could also be done by reconnection between the envelope field and new emerging field nearby. As an example, new emerging
flux can reconnect with the overlying field of a filament
and destabilize it \cite{Feynman_etal1995} or decreases the tension 
of the overlying field so much that the FR starts to erupt \cite{Chen_etal2000}. Again, the relative orientation 
of the two interacting flux systems is important for efficient reconnection to occur leading to ejective eruption. Also, as we have mentioned above, the relative strengths and fluxes can affect the onset of the eruption in this case. Another factor, is how stable the pre-eruptive field is, before the emergence of new flux gets close to it. For instance, if it is close to be unstable and erupt, even a small amount of emerging flux can interact with the pre-eruptive field to trigger its eruption. Therefore, a series of parameters must be taken into account towards understanding the exact role of flux emergence on triggering filament eruptions and CMEs in the close vicinity. Numerical studies regarding the connection between emerging flux and solar eruptions are presented in the next section.  

%%%%%%%%%%%%%%% End of first page %%%%%%%%%%%%%%%%%%%%%

\maketitle

\section{Numerical Simulations}
\label{sec:eruption}

\subsection{Atmospheric MFR formation}

% ============ FIGURE ============================
\begin{figure*}
\centering\includegraphics[width=\textwidth]{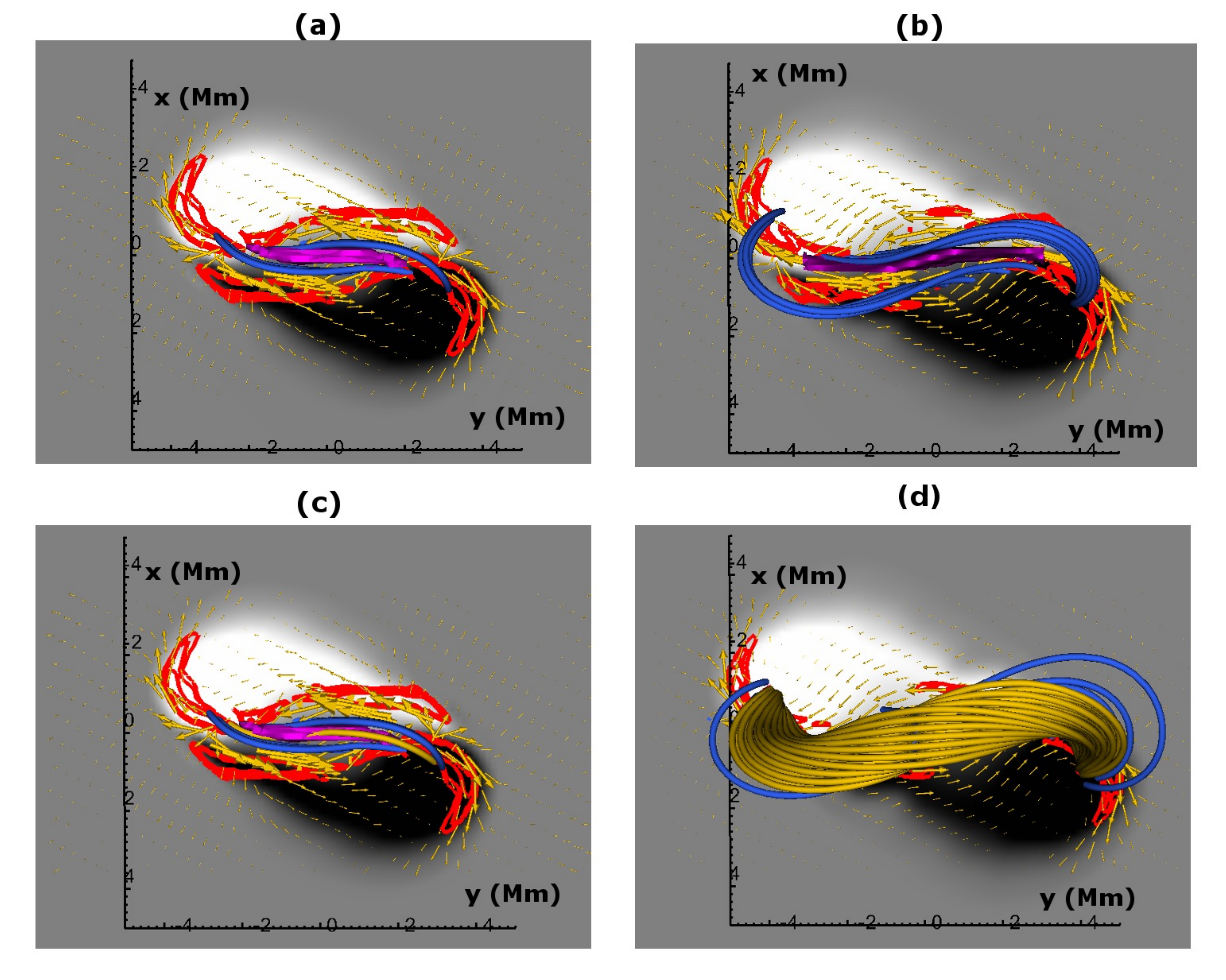}
\caption{ (a, c) Field lines showing the ``sheared arcade'' configuration (blue lines) and the formation of new long magnetic loops (e.g. orange line, c). (b, d)  Field lines showing the J-loops configuration (blue lines) and the formation of a twisted MFR (orange lines, d).  The horizontal slice is photospheric $B_{z}$ (black and white). The yellow arrows show the photospheric velocity field and the red contours show the photospheric vorticity. The purple isosurface is $\left| J/B \right|$. (Figure adapted from Syntelis et al 2017\cite{Syntelis_etal2017}).
}
\label{fig:fig_fr_formation}
\end{figure*}
% ============ FIGURE ============================

A common misconception concerning flux-emergence models is that the atmospheric MFR found in flux-emergence simulations are formed because the whole sub-photospheric flux tube emerges bodily above the photosphere. However, the vast majority of the numerical simulations show that the axis of the emerging flux tube remains below the photosphere or reaches a few pressure scale heights above the photosphere\cite{Fan_2001,Murray_etal2006,Toriumi_etal2013,Syntelis_etal2019b}, as it is very heavy to emerge above the surface and into the atmosphere to form a potentially eruptive MFR. 
Instead, only the field located higher than the axis of the sub-photospheric flux tube emerges and expands into the atmosphere\cite{Archontis_etal2004,Toriumi_etal2011}. This is commonly referred to as \textit{partial emergence} of the sub-photospheric flux tube. During the partial emergence of the flux tube, magnetic polarities are formed at the photosphere and magnetic field expands above the photosphere and into atmosphere. 
The configuration of the magnetic field at and above the photosphere (e.g. bipolar or quadrupolar) will depend on the properties of the sub-photospheric magnetic field\cite{Archontis_etal2014,Fang_etal2014,Syntelis_etal2015,Lee_etal2015,Toriumi_etal2017}.
The gradual emergence of the field at and above the solar surface leads to the self-consistent development of forces, which drive photospheric shearing and converging motions along a PIL\cite{Fan_2001,Manchester_2001} and rotation of the polarities\cite{Fan_2009,Sturrock_etal2015,Sturrock_etal2016}. In the former case, the motions are driven by the Lorenz force parallel to the PIL. 
In the latter case, rotational motions are driven by a torsional \alfven wave propagating from the solar interior to the solar atmosphere, which tends to equate the twist of the field lines in the two regions\cite{Longcope_etal2000}. The combination of these motions, inject shear into the atmosphere.

The continuous shearing of the magnetic field above the photosphere gradually builds a current sheet above the PIL (purple isosurface, Fig.~\ref{fig:fig_fr_formation}a). The sheared magnetic field (blue lines) start to reconnect, forming initially a new flux systems, which consists of long fieldlines that connect the two main polarities above the photosphere (e.g. orange line, panel c). Over time, more shear is injected into the atmosphere. The sheared magnetic field lines gradually adopt a J-loop shape (blue lines, panel b). The reconnection of J-loop field lines form new weakly twisted field lines, forming a sigmoidal twisted MFR (e.g. orange lines, panel d)\cite{Archontis_etal2009,Syntelis_etal2017}. This is a post-emergence MFR and is not associated with the axis of the sub-photospheric flux tube. Instead, the post-emergence MFR is rather formed by a mechanism closely related to the field line connectivities occurring in the internal tether-cutting reconnection\cite{Moore_etal2001} and the ``flux cancellation'' reconnection\cite{vanBallegooijen_etal1989}.
The above process is usually the formation process of atmospheric MFR in flux-emergence simulations \cite{Manchester_etal2004, Fan_2009, Archontis_Torok2008, Archontis_etal2009, Archontis_etal2012, Leake_etal2013a, Leake_etal2014, Syntelis_etal2017}. 
% Such a MFR can become eruptive if an ideal or non-ideal process destabilize it.

% ============ FIGURE ============================
\begin{figure*}
\centering\includegraphics[width=\textwidth]{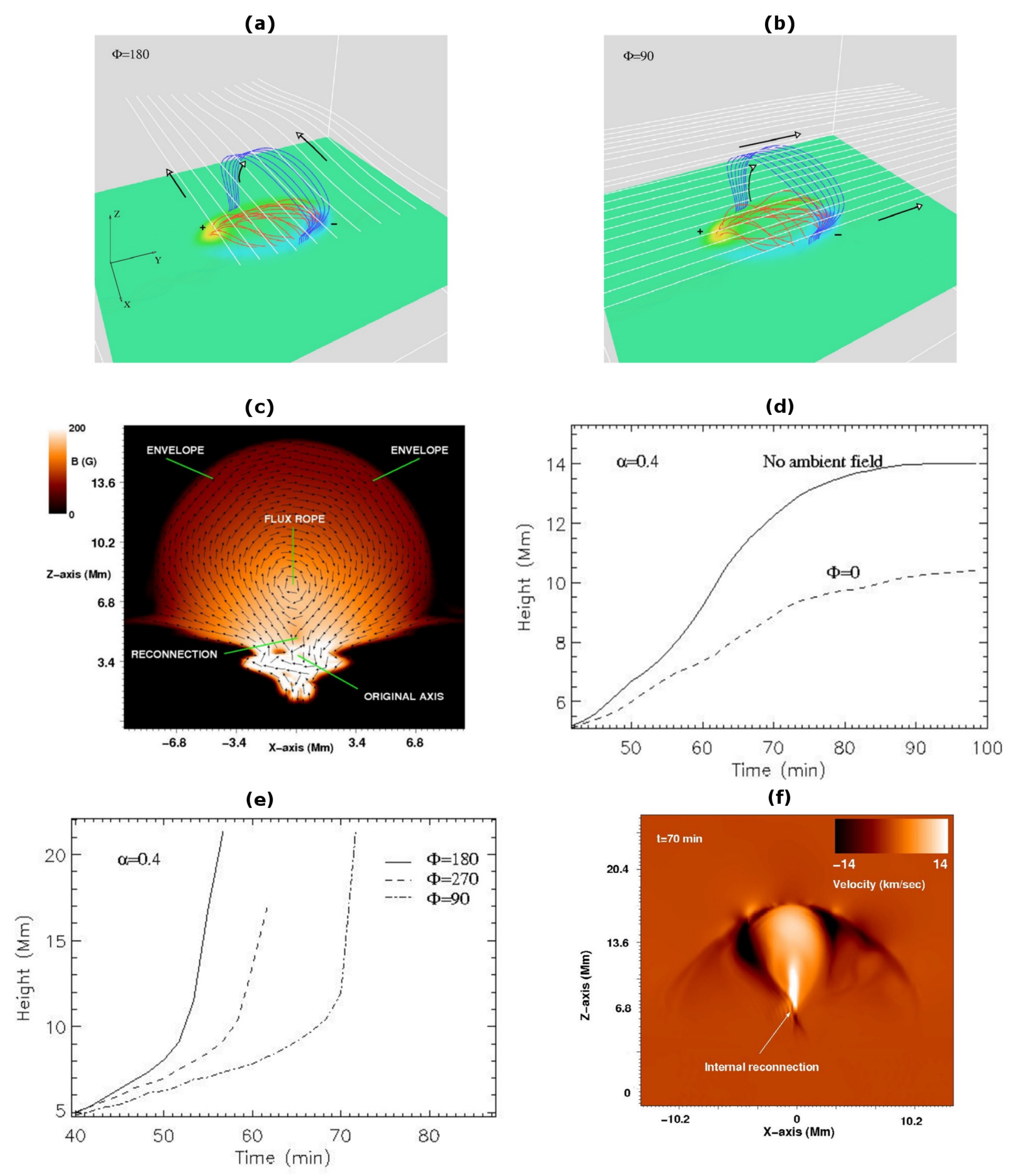}
\caption{ 
(a) Field lines showing the emerged magnetic field and the external magnetic field. The emerged field consists of a core field (similar to the red lines) and an outermost ``envelope'' field (similar to the blue lines). The external field is horizontal and is antiparallel to the envelope field ($\Phi=180\degrees$)  (white lines). 
(b) Same as (a), but for an external field oriented by $\Phi=90\degrees$. 
(c) The magnetic field strength and projected vector at the cross section of the emerged magnetic field region. This panel shows the location of the internal reconnection region, the MFR center, the envelope field and the height of the axis of the sub-photospheric flux tube.
(d) Height-time profile of the MFR for ``confined'' eruptions. The solid line is a case with no external field. Tha dashed line is a case with external field parallel to the envelope field ($\Phi=0$).
(e)  Height-time profiles of the MFR for ``ejective'' eruptions. Different lines show different relative orientation of the field.
(f) $V_z$ component of the velocity at the cross section of the MFR showing the enhanced internal reconnection during fast rising eruptive phase. (Panels are from Archontis et al 2012\cite{Archontis_etal2012}).
}
\label{fig:ambient}
\end{figure*}
% ============ FIGURE ============================

\subsection{Eruptions}

% \subsubsection{Non-Ideal Processes}

% External reconnection
A MFR can be destabilized via non-ideal processes (i.e. magnetic reconnection, see Introduction).
One such non-ideal process associated with MFR eruptions is external (or breakout) reconnection.
This reconnection can occur when the magnetic ``envelope'' field associated with the emerged magnetic flux (blue lines, Fig.~\ref{fig:ambient}a) comes into contact with an external ambient atmospheric field (white lines, Fig.~\ref{fig:ambient}a). 
In order for the external reconnection to occur, the relative orientation of the field lines of the two interacting systems is important \cite{Galsgaard_etal2005, Archontis_Torok2008,Archontis_etal2010}. An example of this relative angle is visualized in Fig.~\ref{fig:ambient}a,b. In panel a, the external field is antiparallel ($\Phi=180\degrees$), and in panel b it is perpendicular ($\Phi=90\degrees$) to the envelope field.  This angle will affect i) whether reconnection will occur and ii) the efficiency of the reconnection. 
The external reconnection is closely related to eruptions\cite{Archontis_Torok2008,Archontis_etal2010,Archontis_etal2012}. 
Archontis et al 2012 \cite{Archontis_etal2012} studied a series of flux emergence simulations where they changed the orientation of the external field, ($\Phi=0-270\degrees$) and compared the results with a field free atmosphere simulation. In all cases, a MFR is formed inside the magnetic envelope through shearing and reconnection (e.g. Fig.~\ref{fig:ambient}c). 
In the case of the field free atmosphere, the MFR rises to coronal heights, but remains ``confined'' by the strapping envelope magnetic field (height-time profile, solid line, Fig.~\ref{fig:ambient}d). 
The presence of an atmospheric field oriented so that it does not favour external reconnection (external field parallel to envelope field, $\Phi=0\degrees$), suppressed the rise of the MFR in comparison to the field free case (dashed line, Fig.~\ref{fig:ambient}d). The MFR remained again confined, but was located lower in the atmosphere in comparison to the field free case.

The eruption occurred only when the relative orientation between the external and the envelope field facilitated efficient reconnection. For instance, Fig.~\ref{fig:ambient}e (solid line) shows the height-time profile of the MFR when the external field is antiparallel to the envelope field. The height-time profile shows a slow rise phase, followed by an ejective fast-rise eruptive phase. 
The eruption occurred as follows. In between the external and the envelope field, a current sheet was formed and external reconnection took place. The external reconnection removed magnetic flux from the strapping envelope field. As a result, the downwards magnetic tension force associated with the strapping envelope field reduced, forcing the MFR to move upwards, causing the slow rise phase. While the MFR moved upwards, the magnetic pressure underneath it was reduced. Plasma flows were then induced towards the low pressure region, enhancing the current sheet underneath the rising MFR. Due to that, internal reconnection was enhanced (Fig.~\ref{fig:ambient}f). Eventually, internal reconnection became very efficient, assisting the acceleration of the MFR. The onset of the fast-rise phase of the eruptions was closely related to the efficiency of the internal reconnection.
Notice that for relative orientation different than the antiparallel one ($\Phi=90\degrees$ or $\Phi=270\degrees$), the eruption occurs later than the $\Phi=180\degrees$ case (Fig.~\ref{fig:ambient}e). This is because the external reconnection becomes less efficient for smaller relative orientations. Therefore, the magnetic flux above the MFR is removed slower, leading to a slower rise phase and a delayed eruption. Similar results were found in simulations where the external field was assumed to be an arcade-like field instead of a horizontal straight field\cite{Leake_etal2013a,Leake_etal2014}.

% \subsubsection{Ideal Processes}

% ============ FIGURE ============================
\begin{figure*}
\centering\includegraphics[width=\textwidth]{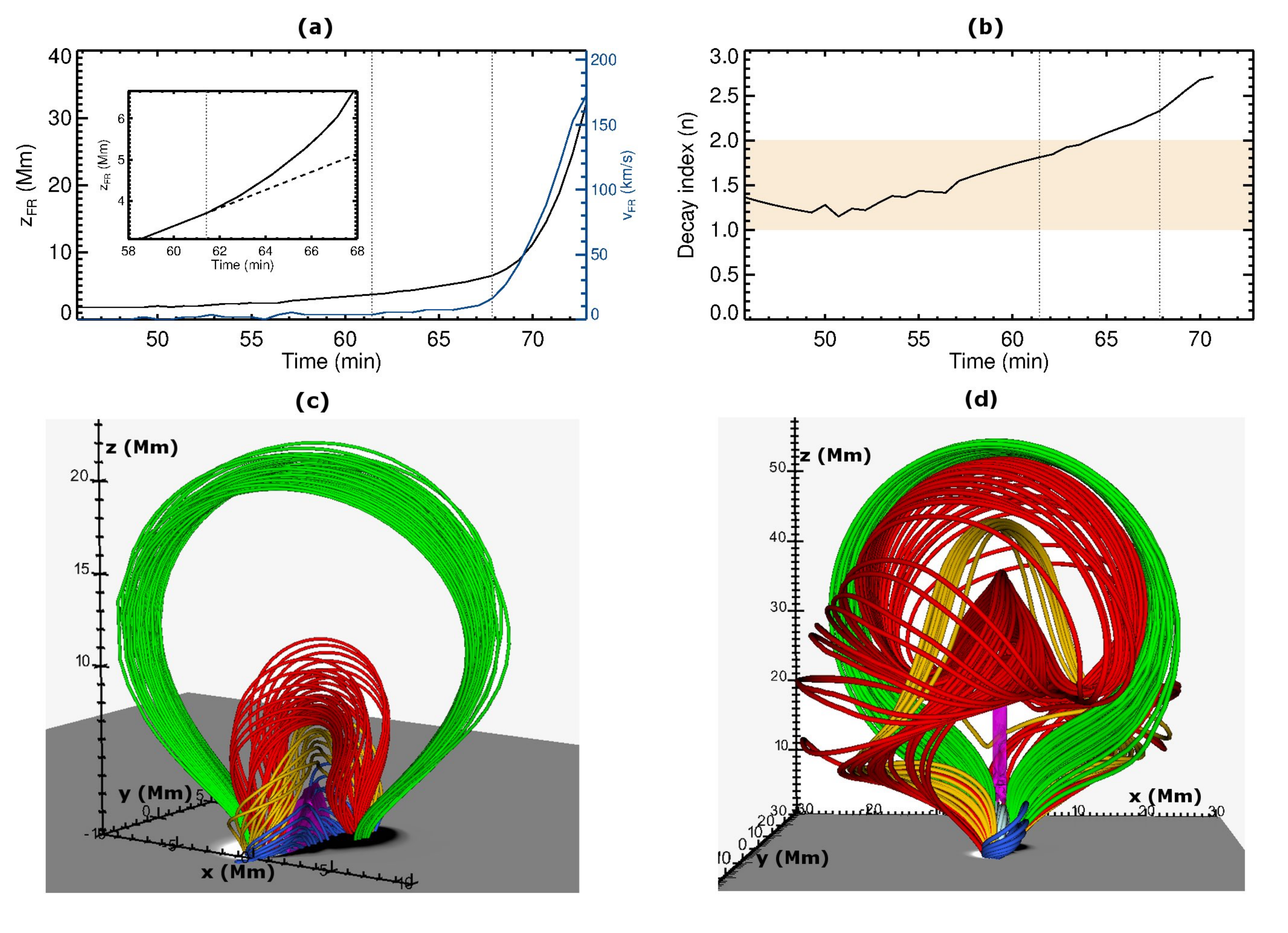}
\caption{ 
(a) Height-time profile of the MFR showing a slow rise and a fast rise phase. The vertical lines indicate the onset of torus instability (left) and tether cutting reconnection (right). The inset shows a close-up of the height–time profile around the initiation of torus instability.
(b) The decay index $n$ measured at the MFR center.
(c) Magnetic field topology prior the tether-cutting reconnection. Blue lines show low lying J-loops. Yellow lines show the MFR. Green lines show the outermost envelope field lines. Red lines show the stretched envelope field lines above the MFR about to reconnect via tether-cutting. The purple isosurface is a low-lying current sheet. 
(d) Same as (c) but after the tether-cutting reconnection. Here, the red lines have reconnected and have now become part of the MFR. Cyan lines show the flare loops (Figure adapted from Syntelis et al 2017\cite{Syntelis_etal2017}).
}
\label{fig:cme-like}
\end{figure*}
% ============ FIGURE ============================

% Torus Inst + tether cutting reconnection
A MFR can be also destabilized also by ideal processes (i.e. MHD instabilities). Such an example was studied using flux-emergence simulations\cite{Archontis_etal2014,Syntelis_etal2017}. In Syntelis et al 2017\cite{Syntelis_etal2017}, a field free atmosphere was assumed in order to avoid any external reconnection. During the simulation, a MFR is formed inside the magnetic envelope through shearing and reconnection (e.g. Fig.~\ref{fig:ambient}c). 
This MFR exhibited a slow rise acceleration phase followed by a fast rise acceleration phase (height-time profile, black line, Fig.~\ref{fig:cme-like}a). 
The slow rise phase was associated with the torus instability (first vertical line, panels a, b), as it coincided with: i) acceleration of the MFR (MFR velocity, blue line, panel a), ii) a high local decay index ($n=1.81$, Fig.~\ref{fig:cme-like}b) and iii) the lack of a strong current sheet underneath the MFR, indicating that the slow rise could not be associated with internal reconnection.

During the slow rise phase, the upwards moving MFR (yellow lines, Fig.~\ref{fig:cme-like}c) pushed and stretched the overlying strapping envelope field (red lines). Due to that, the magnetic pressure underneath the MFR reduced, inducing flows towards the low pressure region. 
Eventually, the stretched envelope field lines reached the vicinity of the (weak) current sheet underneath the rising MFR (red lines bending towards the purple isosurface, panel c). There, the stretched envelope field lines reconnected with other such lines in a tether-cutting manner. This reconnection coincided with the transition from the slow-rise phase to the fast-rise eruptive phase (second vertical line, panel a), triggering the fast ejective eruption. 
During the eruption, a long flare current sheet (purple isosurface, Fig.~\ref{fig:cme-like}d) was formed between the erupting MFR (red and yellow lines) and the flare loops (cyan lines). 
Because of the internal tether-cutting reconnection, flux from the strapping magnetic envelope field was converted into flux of the twisted erupting MFR (red tether-cut lines in panel d now twist around the MFR axis). 
Due to that: i) the MFR grew in physical size and flux, ii) its twist increased dynamically during the eruption and iii) the tension of the strapping envelope field was released. The latter is very important for the eruption. The release of the overlying tension further accelerated the MFR, stretching even further the envelope field, leading to increased rate of tether-cutting reconnection at the flare current sheet. This was a runaway reconnection, that eventually removed all the overlying envelope field and led to a fast ejective CME-like eruption. 
Such eruptions occurred in a recurrent manner. As long as shearing and converging motions are present and magnetic energy is available, post-emergence MFRs can form (and eject) in a recurrent manner \cite{Moreno-Insertis_etal2013,Archontis_etal2014, Syntelis_etal2017}. 
It is important to note that in flux emergence models, \textit{the formation of MFRs and their eruptions does not occur only during the flux emergence phase} (i.e. when photospehric flux increases). 
On the contrary, these can occur also when the photospheric flux has saturated, but the photospheric motions are still active.
Syntelis et al 2019 \cite{Syntelis_etal2019} performed a parametric study by increasing the magnetic field strength of the the sub-photospheric flux tube that led to the recurrent CME-like eruptions, and found recurrent ejective eruptions with higher energies. The energy of the eruptions and the photospheric flux were found to scale linearly in a logarithmic plot.

An interesting result of the Syntelis et al 2017 \cite{Syntelis_etal2017} study was that runaway tether-cutting reconnection occurred in two different ways, i.e. when: i) envelope field lines reconnected with other envelope field lines (envelope-envelope tether-cutting, as described before), and when ii) envelope field lines reconnected with J-loops similar to the blue lines in panel c. It was found that both tether-cutting reconnections acted as tension removal mechanisms and assisted the eruption of the MFRs. 
However, the different location of the tether-cutting had implications for the transfer of hot plasma inside the erupting core. It was found that during the envelope-envelope tether-cutting, hot material from the flare current sheet was pushed towards the center of the erupting structure, forming the typical ``hot'' core found in some CMEs. However, during the envelope-J tether-cutting, the hot material from the flare CS did not reach the center of the erupting core, leaving a mostly ``cold'' core. This is a new mechanism to explain why some CME cores are hot whereas others remain cold \cite{Nindos_etal2015}.

The Syntelis et al 2017\cite{Syntelis_etal2017} study indicated that torus instability did not drive the ejective phase of the CME-like eruption. The torus instability instead accelerated the MFR to the point that the MFR stretched the envelope field enough to trigger runaway tether-cutting reconnection at the flare current sheet. This implied that torus instability alone, without a resistive process, might not be enough to drive a fast ejective eruption.

% ============ FIGURE ============================
\begin{figure*}
\centering\includegraphics[width=\textwidth]{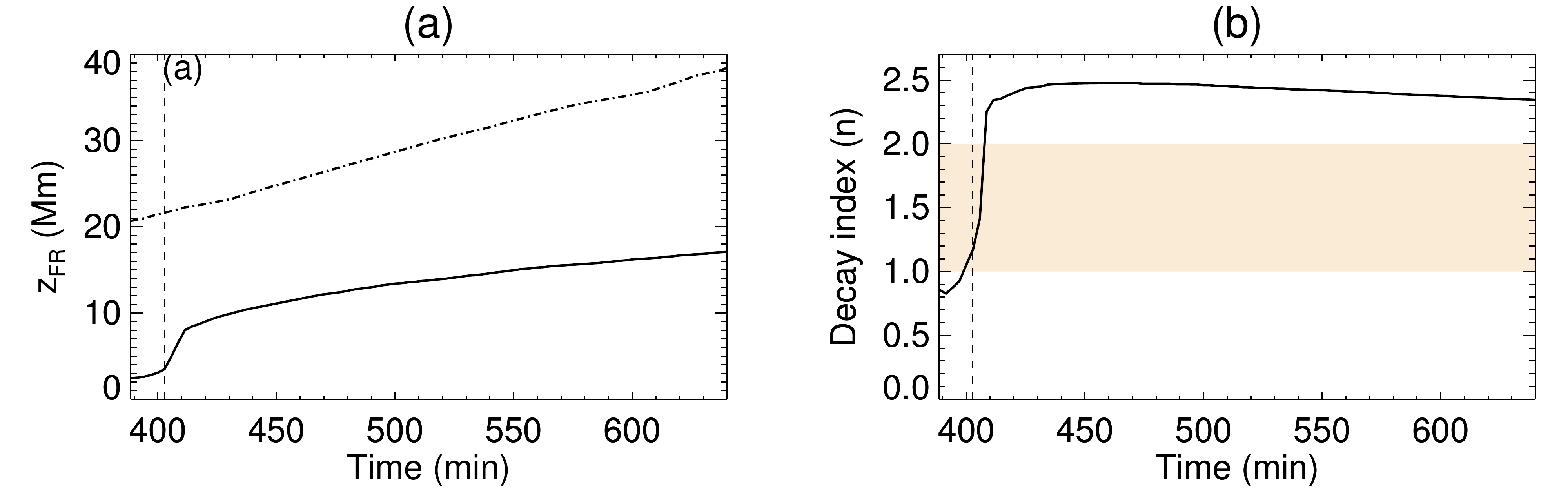}
\caption{ 
(a) Height-time profile of a confined eruption (solid line) inside a high decay index environment. Dashed line shows the height-time profile of the envelope field, inside which the confined eruption takes place.
(b) The decay index measured at the MFR center. 
}
\label{fig:partII}
\end{figure*}
% ============ FIGURE ============================

% Part II forces analysis
We have performed a further (preliminary) study of the above proposition by reducing the magnetic field strength of the sub-photospheric flux tube that led to the recurrent CME-like eruptions. The result was a non-ejective eruption (height-time profile, solid line Fig.~\ref{fig:partII}a) inside a high decay index envelope field (Fig.~\ref{fig:partII}b). 
To understand the dynamics of the confined eruption, we focused on the analysis of the forces such as: i) the upwards directed hoop force, ii) the downwards directed self-tension force (i.e. the force occurring from the poloidal current of the MFR and the toroidal component of MFR's magnetic field) and iii) the downwards directed ``strapping'' force of the envelope field. This analysis was inspired by the laboratory plasma experiments of erupting MFRs \cite{Myers_etal2015,Myers_etal2016}, which revealed that a MFR eruption is stopped by the combined action of both the self-tension and the ``strapping'' force against the hoop force. Our preliminary results show a similar behaviour, indicating that the self-tension is an important force acting against the development of torus instability.

We should highlight that the photospheric shearing along the polarity inversion line, which develops self-consistently at these models, is not strong enough to twist the magnetic field to the extend to form kink unstable, pre-eruptive MFRs. Thus, the role of kink instability on the eruptions of MFRs in this type of flux emergence models has not been studied yet.

\subsection{Eruption Driven Solar Jets}

% ============ FIGURE ============================
\begin{figure*}
\centering\includegraphics[width=\textwidth]{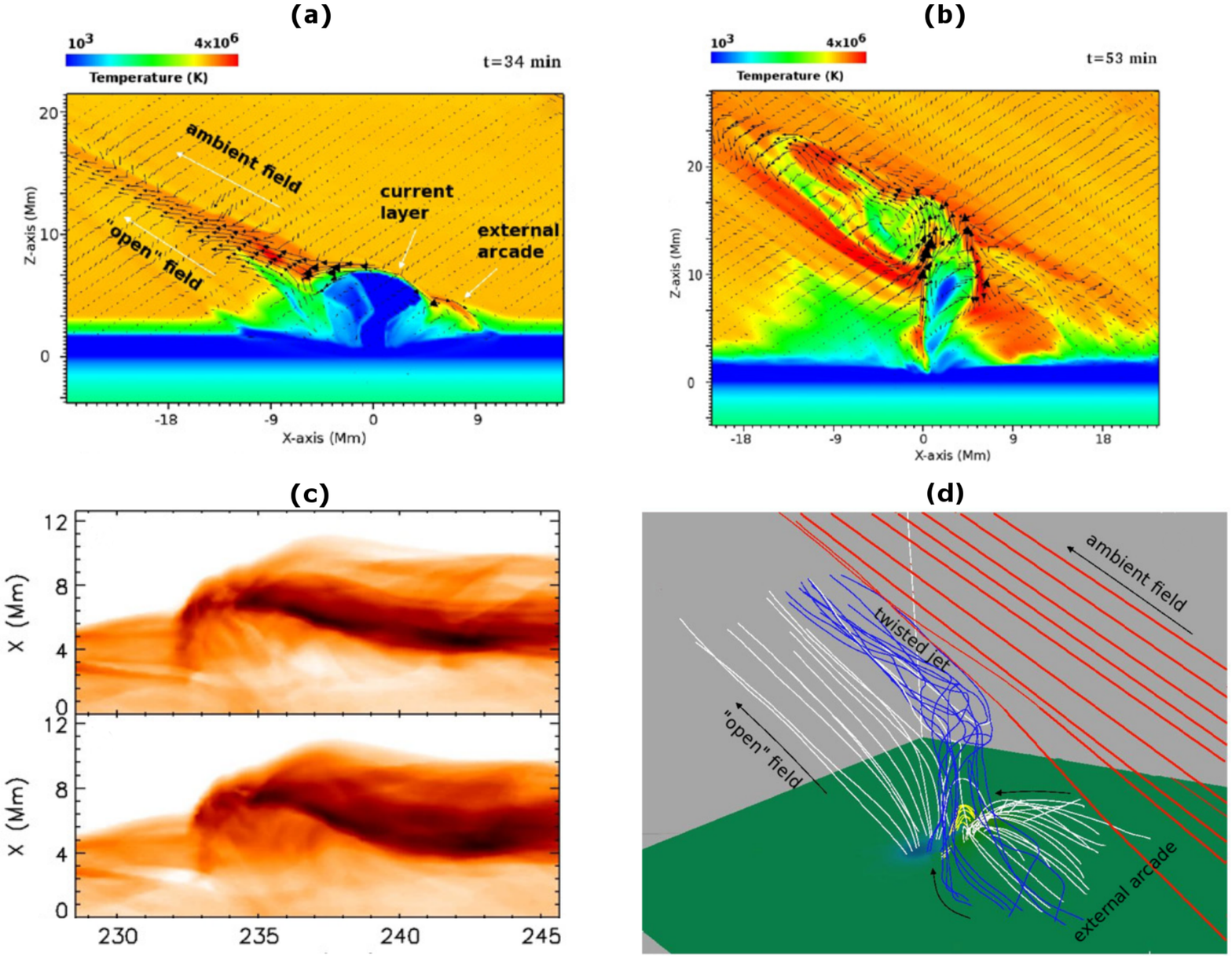}
\caption{ 
(a) Temperature distribution at the vertical midplane showing the configuration prior to the blowout jet eruption. Arrows indicate the projected velocity field on the plane. 
(b) Temperature distribution during the MFR eruption, leading to the onset of  the blowout jet.
(c) Distance-time diagram of $\rho^2$, showing the propagation of the \alfven wave along the spire of a blowout jet.
(d) 3D magnetic field topology during the ejection of the blow-out jet. See text for details (Panels a,b,d are adapted from Archontis and Hood (2013) \cite{Archontis_etal2013b}. Panel c is adapted from Lee et al (2015) \cite{Lee_etal2015}).
}
\label{fig:jets1}
\end{figure*}
% ============ FIGURE ============================

Eruptions of small-scale MFRs (called minifilaments) have been associated with the so-called ``blowout'' jets.
Blowout jets have been extensively studied using flux-emergence simulations\cite{Archontis_etal2013b, Fang_etal2014,Lee_etal2015,Moreno-Insertis_etal2013}. The eruption of a blowout jet was studied by Archontis \& Hood 2013 \cite{Archontis_etal2013b}, focusing on the interaction between an emerged magnetic region inside and an oblique external ambient magnetic field (Fig.~\ref{fig:jets1}a). During the interaction of the systems, a current sheet was formed between the emerged field and the external field (``current layer''). There, external reconnection initially formed some new ``open'' field lines that carried a hot narrow bi-directional flow, and a new external arcade field adjacent to the emerged magnetic envelope. Inside the magnetic envelope (blue region underneath ``current layer''), a MFR was formed above the PIL through shearing and reconnection.
The external reconnection removed flux and tension from the envelope field. This made the MFR to slowly rise upwards, reducing the magnetic pressure underneath the MFR, inducing inflows towards the current sheet underneath the MFR. At that current sheet, slow internal reconnection took place, increasing the MFR's flux and size. Eventually the MFR erupted outwards due to both external and internal reconnection. During the eruption, the MFR reconnected with the straight ambient field. The erupting field and the associated cool material were channeled along the straight external field (Fig.~\ref{fig:jets1}b). 
The reconnection of the twisted MFR with the untwisted ambient field created an untwisting motion, forming a ``rotating'' jet. This untwisting motion was developed due to the propagation of a torsional \alfven wave along the open field \cite{Lee_etal2015} (Fig.~\ref{fig:jets1}c). The above process eventually created a twisted and wide blowout jet, that carried upwards both cool and hot plasma (Fig.~\ref{fig:jets1}d). 
Models focusing on the shearing along a PIL show similar results, whereby the eruption of a MFR is triggered by the combined action of external and internal reconnection, leading to the ejection of a twisted blowout jet \cite{Wyper_etal2017,Wyper_etal2018}.

% ============ FIGURE ============================
\begin{figure*}
\centering\includegraphics[width=\textwidth]{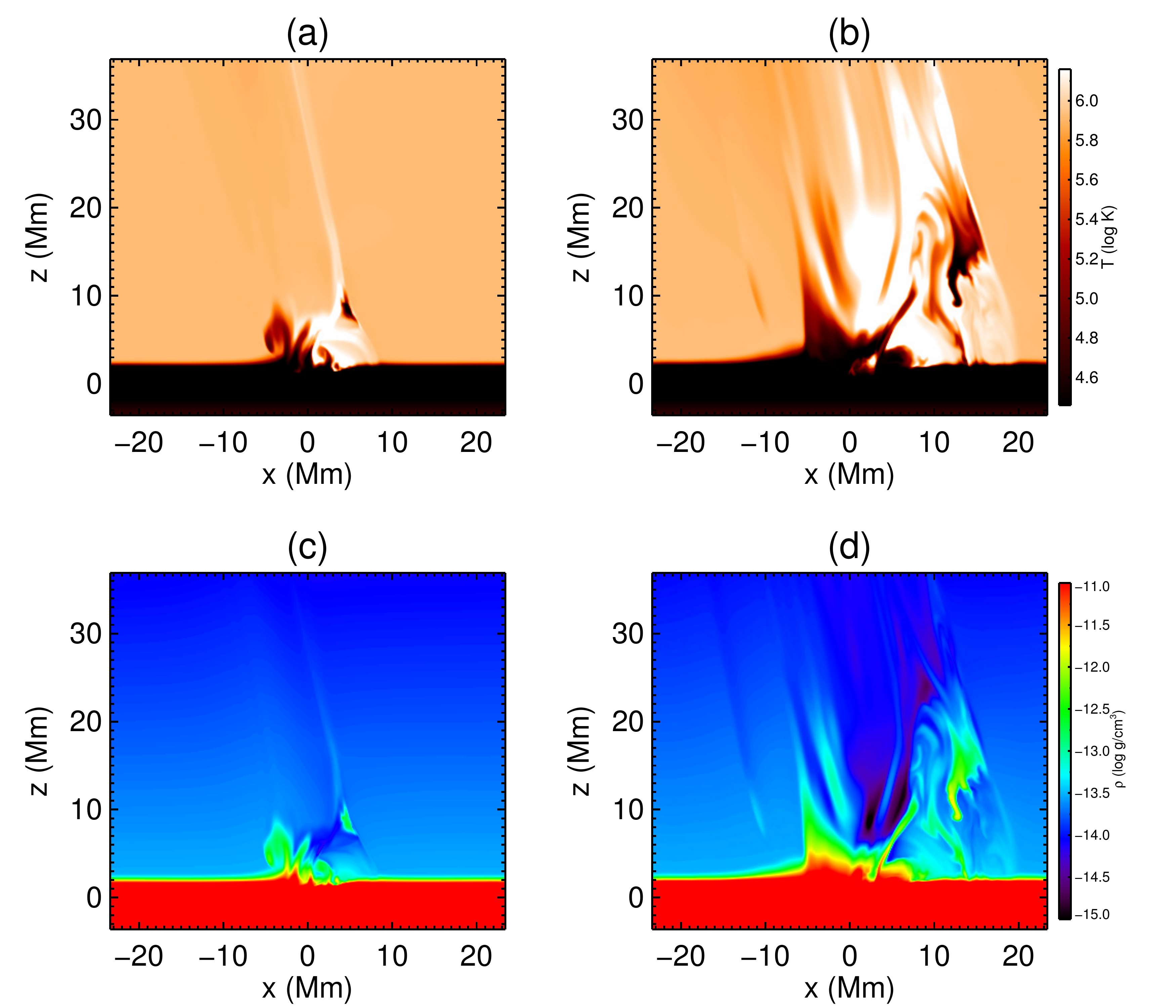}
\caption{ Temperature (first row) and density (second row) of a standard (first column) and a blowout (second column) jet driven by a minifilament eruption.
}
\label{fig:jets2}
\end{figure*}
% ============ FIGURE ============================

Recent observations have shown that minifilament eruptions are associated not only with blowout jets, but also with standard jets\cite{Sterling_etal2015,Moore_etal2018}. 
We examined numerically the hypothesis that minifilament eruptions drive all coronal jets, from standard to blowout.
Preliminary results show that this can indeed happen. 
Fig.~\ref{fig:jets2}a,c shows the temperature and density of a standard jet driven by a minifilament eruption. In this case, hot material flows along  the narrow spire of the jet. At the base of the spire, the cool and dense material is due to an erupting minifilament. This cool material does not propagate along the narrow spire. Instead, it is found to remain at lower atmospheric heights.
On the other hand, in Fig.~\ref{fig:jets2}b,d a more energetic and larger minifilament drives a blowout jet. There, the width of the spire is comparable to the size of the base of the jet. Both hot and cool material is ejected along the spire of the jet.

\section{Summary}

In this review, we have discussed the role of magnetic flux emergence on the onset of some of the most dynamic solar phenomena. Observations and numerical studies highlight that the process of flux emergence is crucial, towards understanding the formation of ARs, which are the building blocks of the solar activity. Also, they show that flux emergence is a multi-scale process, which can be responsible for the magnetic and thermo-dynamic coupling of the highly stratified solar atmosphere. In addition, it can drive and/or trigger powerful eruptions (e.g. blowout jets, CMEs), especially in conjunction with other important physical processes, such as shearing and reconnection of magnetic fieldlines.

Considerable progress has been made in exploring the exact mechanism(s) leading to solar activity (e.g. eruptions, jets) in emerging flux regions during, both, the early and later stage of emergence. For instance, in numerical simulations of flux emergence, it has been reported that recurrent ejective eruptions of MFRs is a combination of an ideal instability (torus instability) and a non-ideal process (tether cutting reconnection of the envelope field). These eruptions may drive the onset of blowout jets or evolve into CMEs, releasing a considerable amount of energy and flux towards the outer space.

Further advancements in parallel computing will help us to study the formation of ARs, taking into account: (i) the interplay between flux emergence and realistic magneto-convection and (ii) the effect of global convective dynamo action on the formation of magnetic fields, which can emerge at the solar surface forming pores/sunspots. Then, one has to follow the long-term evolution of the emerging field, in order to study the response of the stratified solar atmosphere and the onset of eruptive events, as we mentioned above. Obviously, high-resolution observations (e.g. photospheric magnetograms) are needed to study the process of flux emergence at various scales, from the small scale internetwork field to the largest ARs. Helioseismic observations can also provide us with data and important information regarding the emergence of magnetic fields below the photosphere by e.g. measuring the flow patterns associated with strong sub-surface magnetic flux concentrations of sunspot and emerging ARs. Therefore, the combination of  observations of emerging flux at the photosphere, helioseismology and advanced 3D simulations is the next step towards a better understanding of how dynamo-generated magnetic fields are related to the evolution of emerging flux at the solar surface and the role of magnetic flux emergence on the onset of solar dynamic events.

%%%%%%%%%% Ethics etc %%%%%%%%%%%%%%
\enlargethispage{20pt}
% \ethics{Insert ethics statement here if applicable.}
\dataccess{The article has no additional data.}
\aucontribute{Both authors drafted, read and approved the manuscript'.}
\competing{The author(s) declare that they have no competing interests.}
\funding{VA is supported by the Royal Society, through a University Research Fellowship. PS is supported by an STFC consolidated grant.}
%\ack{Insert acknowledgment text here.}
% \disclaimer{Insert disclaimer text here if applicable.}

%%%%%%%%%% Insert bibliography here %%%%%%%%%%%%%%

\bibliographystyle{rsta}
\bibliography{bibliography}

\end{document}